\documentclass[11pt,twoside]{article}
\usepackage{asp2004}
\usepackage{epsf}
\usepackage{graphics}
\usepackage{amsmath,epsfig,rotating,amssymb}
\usepackage{lscape}
\def\edcomment#1{\iffalse\marginpar{\raggedright\sl#1\/}\else\relax\fi}
\reversemarginpar

\begin{document}
\title{Influence of the magnetic field on the thermal condensation}
 \author{P. Hennebelle$^1$ and T. Passot$^2$}
\affil{$^1$ Laboratoire de radioastronomie millim{\'e}trique, UMR 8112 du CNRS, 
\newline {\'E}cole normale sup{\'e}rieure et Observatoire de Paris, 24 rue Lhomond,
\newline 75231 Paris cedex 05, France  
\newline $^2$ CNRS, Observatoire de la C\^ote d'Azur, B.P.\ 4229, 06304, Nice
 Cedex 4, France}

\begin{abstract}
The neutral atomic interstellar medium (HI) is well known to be 
strongly magnetized. Since HI is a 2-phase medium the questions arise 
of what is the effect of the magnetic field on a 2-phase medium, how
magnetic field affects thermal instability ? 
Here we summarize analytical and numerical results which have been 
obtained previously to investigate this question.
\end{abstract}

\vspace{-0.5cm}
\section{Introduction}

\begin{figure} [!ht]
\includegraphics[width=8cm]{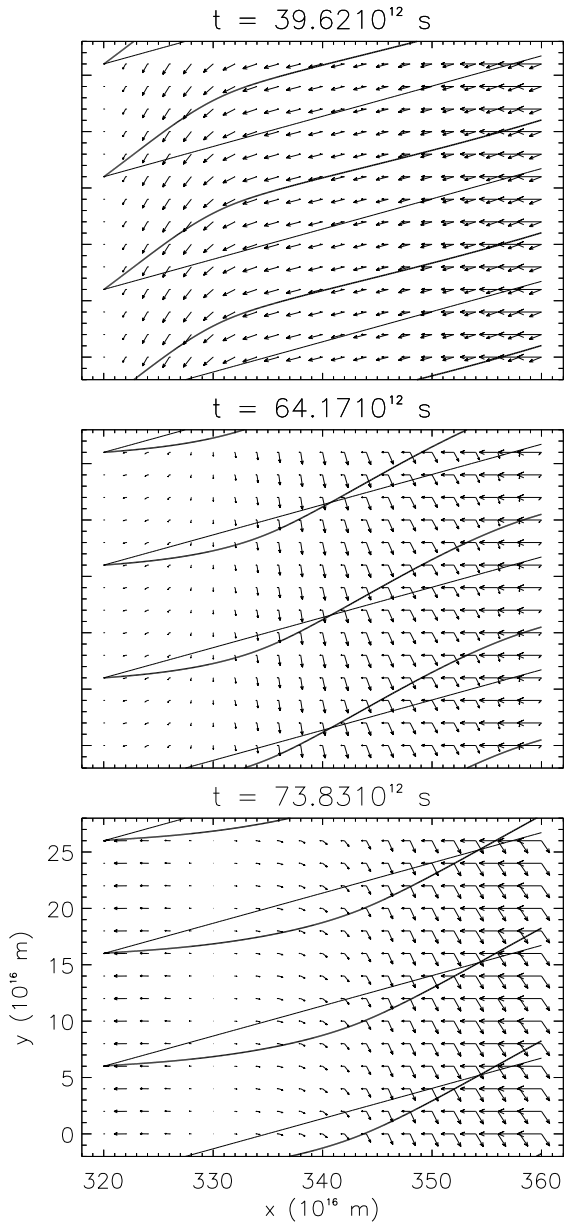}
\caption{Velocity (arrows) and magnetic field lines (thick solid lines)
at 3 time steps. Only half of the solution is displayed since it is symmetrical
with respect to the y-axis.}
\label{fig1}
\end{figure}

\begin{figure}[!ht]
\includegraphics[width=8cm]{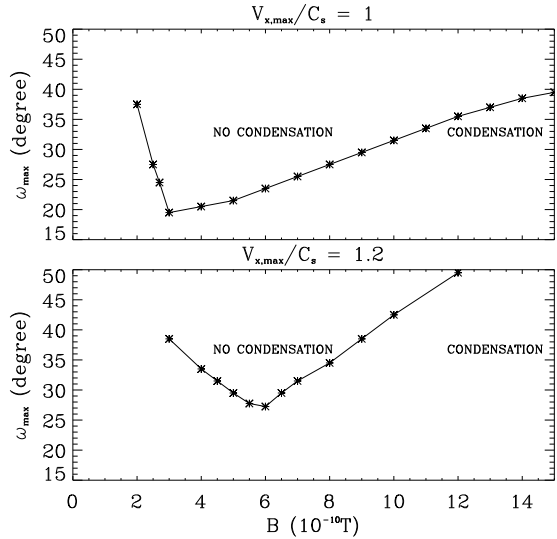}
\caption{Condensation threshold as a function of magnetic 
intensity and of the initial angle 
between velocity and magnetic field, $\omega$.}
\label{fig2}
\end{figure}

\begin{figure}[!ht]
\plottwo{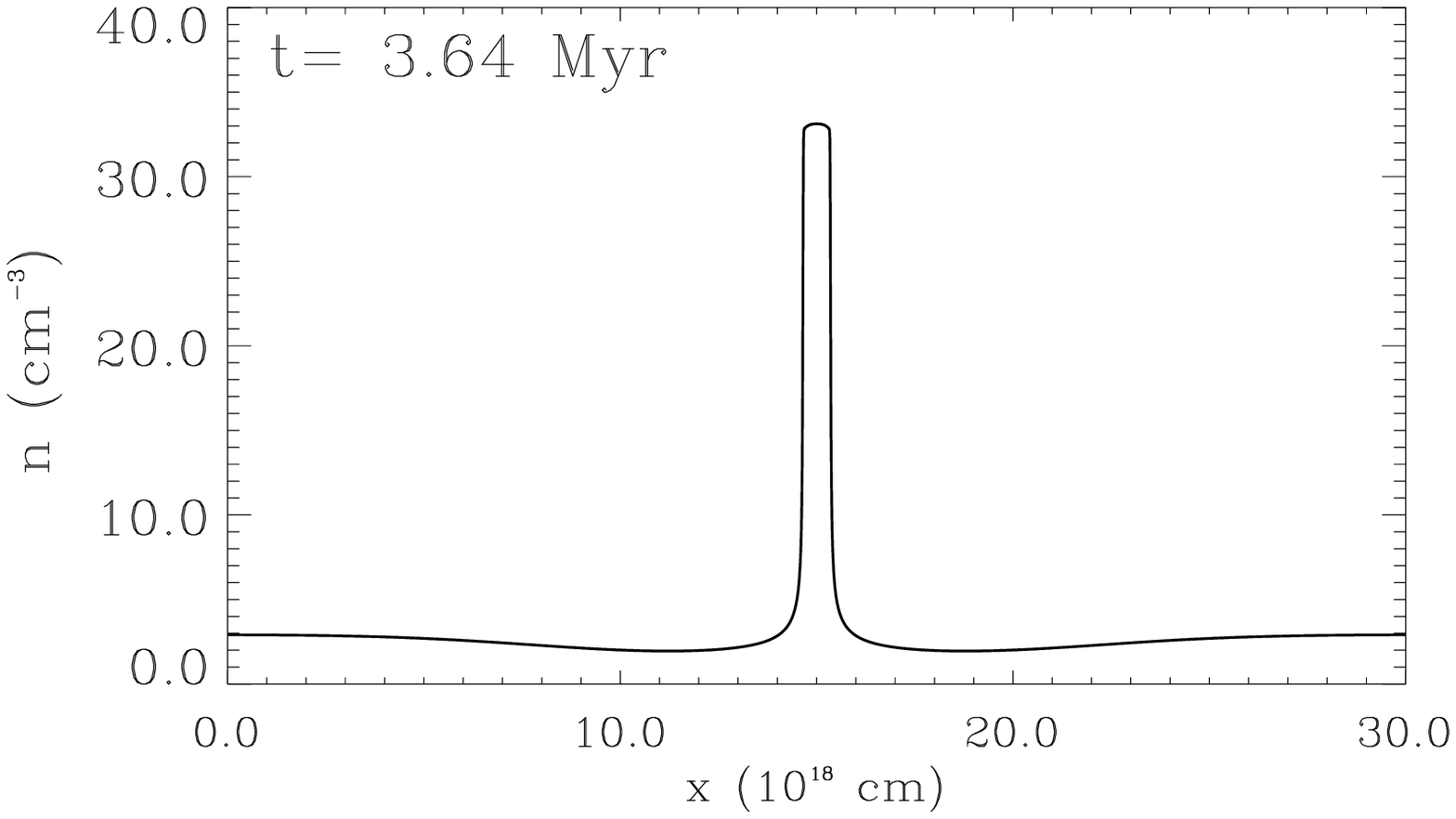} {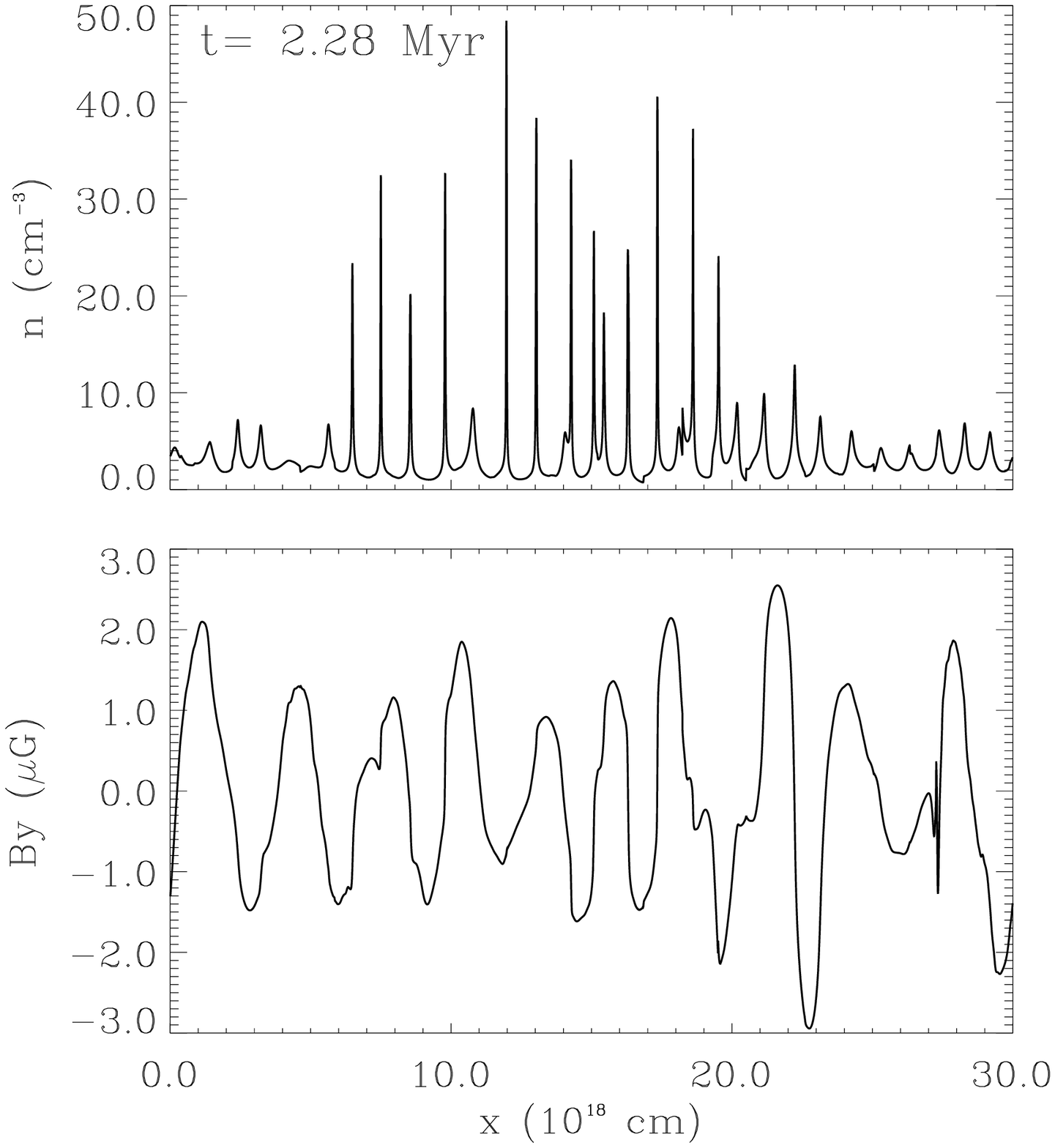}
\caption{Left panel: CNM structure formed through the evolution
of a density perturbation in thermally unstable gas. Right panels:
Density field and magnetic field y-component when an Alfv\'en 
wave is setup in initially.}
\label{fig3}
\end{figure}

Measurements of magnetic intensity in the interstellar atomic 
hydrogen reveal that the magnetic intensity is about 
6$\mu$G (Troland \& Heiles 1986, Heiles 1987, Heiles \& Troland 2005). 
This indicates that the  magnetic energy overcomes the 
thermal energy by a factor of a few.  
Since the density in HI varies over 
at least 2 orders of magnitude, the question arises of how the gas 
can condense so strongly in spite of a strong magnetic field without 
the help of gravity which is negligible in HI.
Moreover, the magnetic intensity is observed to be nearly 
independent of the gas density. More generally knowing the effect 
of the magnetic field on the development of the thermal instability 
is of great relevance for HI. 
Here we present some results obtained either analytically or by the mean
of 1D numerical simulations to answer these questions.  
In the second section, we consider the case of an initially 
uniform  magnetic field and identify precisely a mechanism 
which permits naturally the condensation to proceed along the field lines.
In  section 3, we investigate the effects of circularly 
polarized Alfv\'en waves on thermal instability. Section 4 concludes
the paper.

\section{Effect of an initially uniform magnetic field}
Dynamical motions such as shocks (Koyama \& Inutsuka 2000) or converging flows
  (hennebelle \& P\'erault 1999) have been proposed to be an efficient 
way of triggering thermal instability.  Here we study how an initially
uniform magnetic field making an angle $\omega$ with the initial 
velocity field affects the condensation process. 
Figure 1 shows the results of a 1D numerical simulation. 
The configuration being symmetrical with respect to y-axis, only 
half of the solution is displayed. The arrows 
display the velocity field whereas the thick solid line is the magnetic 
field line at the current time step. The thin solid line is the initial 
magnetic field line. Three timesteps are displayed. 
First panel shows that the converging flow has bended the field lines
therefore increasing the magnetic pressure. This has the effect to slow
down the gas and tends to stop the condensation. However in the same way, 
magnetic tension has generated a transverse flow. 
Second panel reveals that the transverse flow has significantly unbended  
the magnetic field lines and therefore decreases the magnetic pressure.
As a consequence the condensation can proceed along the field line as 
shown in third panel. Figure 2 shows the condensation threshold for various
values of  the angle $\omega$ and the magnetic intensity for 2 different 
values of the initial velocity field. When the magnetic intensity is small, 
$\omega$ decreases rapidly when $B$ increases. This is a natural consequence of
the magnetic pressure being stronger. However for intermediate values of $B$
($\simeq 3-5 \mu$G), $\omega$ does not decreases anymore when $B$ increases 
and for strong values of $B$, $\omega$ increases with $B$. This behaviour is a 
direct consequence of the magnetic tension which tends to unbend the field 
lines. Indeed when the field is strong so that the magnetic energy
dominates the kinematic one, the gas is constraint to flow along the field 
lines. 

An important consequence of this mechanism is that no correlation between the
magnetic intensity and the density is expected in HI. This result first 
obtained in 1D by Hennebelle \& P\'erault (2000) 
(see also Passot \& V\'azquez-Semadeni 2003) has been confirmed 
in 3D simulations by de Avillez \& Breitschwerdt (2005).

\section{Effect  the magnetic waves}
Since magnetic and kinematical energies are roughly comparable in the HI, 
one expects  magnetic field fluctuations to be important. We have therefore
explored analytically and numerically the influence of Alfv\'en waves on 
 thermal instability (Hennebelle \& Passot 2006). In particular 
an exact stability 
analysis of a non-linear circularly polarized Alfv\'en wave, of wavelength 
$\lambda_0$,  propagating in a 
thermally unstable medium has been worked out. The conclusions of this 
analysis are i) wavelengths larger than $\lambda_0$ are generally 
stabilized against thermal instability by the Alfv\'en wave, ii)
wavelengths slightly smaller than $\lambda_0$ are generally destabilized 
and even more prone to thermal instability. This is a consequence 
of the parametric instability studied by Goldstein (1978). iii) 
The wavelengths much smaller than $\lambda_0$ are not affected 
by the Alfv\'en wave. In order to confirm these results and to explore the 
non linear regime, we have performed 1D numerical simulations. We setup
a density perturbation into a thermally unstable gas that we let evolve. 
Left panel of figure 3 shows that the perturbation has condensed and formed a 
CNM structure. When an Alfv\'en wave is added, right panel of figure 3 shows
that the evolution is drastically different. Instead of one, about 10 
small structures of CNM have formed. This is because as shown by the 
analytical calculations, the waves have triggered the growth of smaller 
wavelengths. This mechanism together with turbulence (Audit \& Hennebelle 2005)
 is likely to play an important r\^ole in the 
formation of the small scale CNM structures recently observed by 
Braun \& Kanekar (2005) and Stanimirovi\'c \& Heiles (2005)

\end{document}